\documentclass[reprint,amsmath,amssymb,aps,floatfix]{revtex4-1}
\usepackage[pdftex]{graphicx}
\usepackage{dcolumn}
\usepackage{bm}
\usepackage{algorithm}
\usepackage{algorithmic}

\begin{document}

\title{Search for Common Minima in Joint Optimization of Multiple Cost Functions}

\author{Daiki Adachi$^1$}
\author{Naoto Tsujimoto$^1$}
\author{Ryosuke Akashi$^1$}
\author{Synge Todo$^{1,2,3}$}
\author{Shinji Tsuneyuki$^{1,2}$}
\affiliation{
  $^1$Department of Physics, University of Tokyo, Tokyo, 113-0033, Japan\\
  $^2$Institute for Solid State Physics, University of Tokyo, Chiba, 277-8581, Japan\\
  $^3$Research and Services Division of Materials Data and Integrated System, National Institute for Materials Science, 1-2-1 Sengen, Tsukuba, Ibaraki 305-0047, Japan
}

\begin{abstract}
We present a novel optimization method, named the Combined Optimization Method (COM), for the joint optimization of two or more cost functions.
Unlike the conventional joint optimization schemes, which try to find minima in a weighted sum of cost functions, the COM explores search space for common minima shared by all the cost functions. 
Given a set of multiple cost functions that have qualitatively different distributions of local minima with each other, the proposed method finds the common minima with a high success rate without the help of any metaheuristics.
As a demonstration, we apply the COM to the crystal structure prediction in materials science.
By introducing the concept of data assimilation, i.e., adopting the theoretical potential energy of the crystal and the crystallinity, which characterizes the agreement with the theoretical and experimental X-ray diffraction patterns, as cost functions, 
we show that the correct crystal structures of Si diamond, low quartz, and low cristobalite can be predicted with significantly higher success rates than the previous methods.
\end{abstract}

\maketitle

Continuous optimization, i.e., finding a global minimum of a continuous nonlinear cost function, is one of the most fundamental and important problems encountered in almost all the fields of science and engineering.
For solving the optimization problem, a variety of classical optimization algorithms, such as the gradient descent, conjugate gradient, Newton and quasi-Newton methods, downhill simplex method, etc. have been developed so far and long been used widely~\cite{Fletcher1987}.
If the cost function has a rugged landscape in a high-dimensional search space, however, such classical algorithms easily fail to find the global optimal point, and get trapped by local minima, or metastable states.
In order to overcome the shortcoming in the classical schemes, various types of so-called {\em metaheuristics}, which are higher-level problem-independent frameworks, have been developed~\cite{BianchiDGG2009}.
Typical examples of metaheuristics are; random search~\cite{Rastrigin1963}, simulated annealing~\cite{KirkpatrickGV1983}, genetic algorithm~\cite{Holland1975}, particle swarm optimization~\cite{KennedyE1995,ShiE1998}, exchange Monte Carlo~\cite{HukushimaN1996}, Bayesian optimization~\cite{Mockus1974}, etc.

Theoretical structure prediction is one of the most important applications of such continuous optimization in physics, chemistry, biology, and pharmacy.
Recently, thanks to the development in computational power of the state-of-the-art supercomputers, the metaheuristic approach to the theoretical potential energy of the crystal has also been applied vigorously to the structure prediction for various materials with great success~\cite{PickardN2011, ChenSPSM2016, PannetierACC1990, DollSJ2008, AbrahamP2006, WuJWNZUWH2013, SupadyBB2015, WangLZM2010, WangLZLYLWZM2015, MuggletonKS1992, RaccugliaEAFWMZFSN2016, LeeSSNT2016}.
The prediction ability of such theoretical methods, however, becomes depleted quite rapidly as target materials become more and more complex.
This is due to an exponentially large search space with the increase of the number of atoms, and also to an extremely rugged landscape of the potential energies~\cite{GlassOH2006}.

One of the promising approaches against the exponential complexity is 
to utilize available experimental data for support.
It is expected that, by taking a weighted sum of two or more different cost functions, a part of trivial local minima from each cost function could be removed, and thus the success rate would increase.
Again in the crystal structure prediction, for example, Puts, Sch\"{o}n, and Jansen proposed to use a combined cost function composed of some approximate potential energy and the coincidence with experimental x-ray diffraction data. Some real crystal structures have been reproduced by using this method~\cite{DinnebierHLJ2009, NapolitanoDCPHME2013} much more efficiently than the methods that utilize only the experimental data, such as the direct space method~\cite{HarrisT1996, KeenM1990} and the maximum entropy method~\cite{TremayneLMBHSGB1992}.
More recently, the present authors demonstrated that
the crystal structure prediction with reliable potential energy can be made quite efficient and robust by combining with partial and incomplete experimental data~\cite{TsujimotoAATT2018}.

Optimizing a sum of multiple cost functions is called the {\em Pareto optimization}~\cite{Pareto1896, CoelloB2003}.
In the Pareto optimization, multiple cost functions are summed up to make a new cost function so as to avoid the overfitting or the local minima of each cost function.
In such cases, a fine tuning of the weights between the cost functions is inevitable
thus an inappropriate choice of the weights can easily give wrong answers.

Here, it should be pointed out that, in the application of the joint optimization, the distribution of local minima of the multiple cost functions often have a special property, i.e., all the cost functions share the same minimum as the global minimum.
Indeed, in the example of the crystal structure predictions discussed above, the correct crystal structure is (at least approximately) a global minimum of the theoretical potential energy as well as that of the difference from the experimental data.
This special property makes the Pareto optimization more robust against the choice of the weights between the cost functions~\cite{TsujimotoAATT2018}.
Furthermore, as demonstrated below, if the cost functions have such a property, the global minimum is obtained by searching for a common minimum, instead of the minimum of the combined cost function, with a much higher success rate.
In this paper, we present a new optimization method for the joint optimization, named the Combined Optimization Method (COM), that explores the search space for common minima shared by the multiple cost functions.
We demonstrate the effectiveness of our method by applying it to the crystal structure prediction in materials science by combining the theoretical potential energy of the crystal and the crystallinity, which characterizes the agreement with the theoretical and experimental X-ray diffraction patterns, as the cost functions.
They are expected to have qualitatively different distributions of local minima besides the global minimum, which corresponds to the correct crystal structure.
We apply our method to Si diamond structure, low quartz, and low cristobalite, and show that the correct target structures can be found with significantly higher success rate than the previous work.

\section*{Combined Optimization Method (COM)}

Our algorithm is a non-trivial extension of the {\em gradient descent}.
Let $f({\bf x})$ be the multi-variable cost function to be minimized, where ${\bf x} = (x_1,x_2,\cdots,x_D)$ is a $D$-dimensional vector.
In the standard gradient descent, we try to decrease $f({\bf x})$ by moving ${\bf x}$ in the direction of the negative gradient of $f$ (or {\em force}), i.e.,
\begin{align}
  x_i \leftarrow x_i - \gamma \frac{\partial f}{\partial x_i} ({\bf x}) \qquad \text{for $i=1,\cdots,D$},
\end{align}
where the parameter $\gamma > 0$ is called the learning rate. In the case of joint optimization of $N$ cost functions, $f_1({\bf x}), \cdots, f_N({\bf x})$, a linear combination of the cost functions,
\begin{align}
  f({\bf x}) = \sum_{k=1}^N \alpha_k f_k({\bf x}),
  \label{eqn:joint}
\end{align}
is often introduced, where $\alpha_k$ ($k=1,\cdots,N$) denotes the weight (or the importance) of the $k$-th cost function, $f_k$.
The gradient descent will find one of local minima of the cost function; the result strongly depends on the initial condition.

\begin{figure}[tbp]
  \centering
  \includegraphics[clip, width=\linewidth]{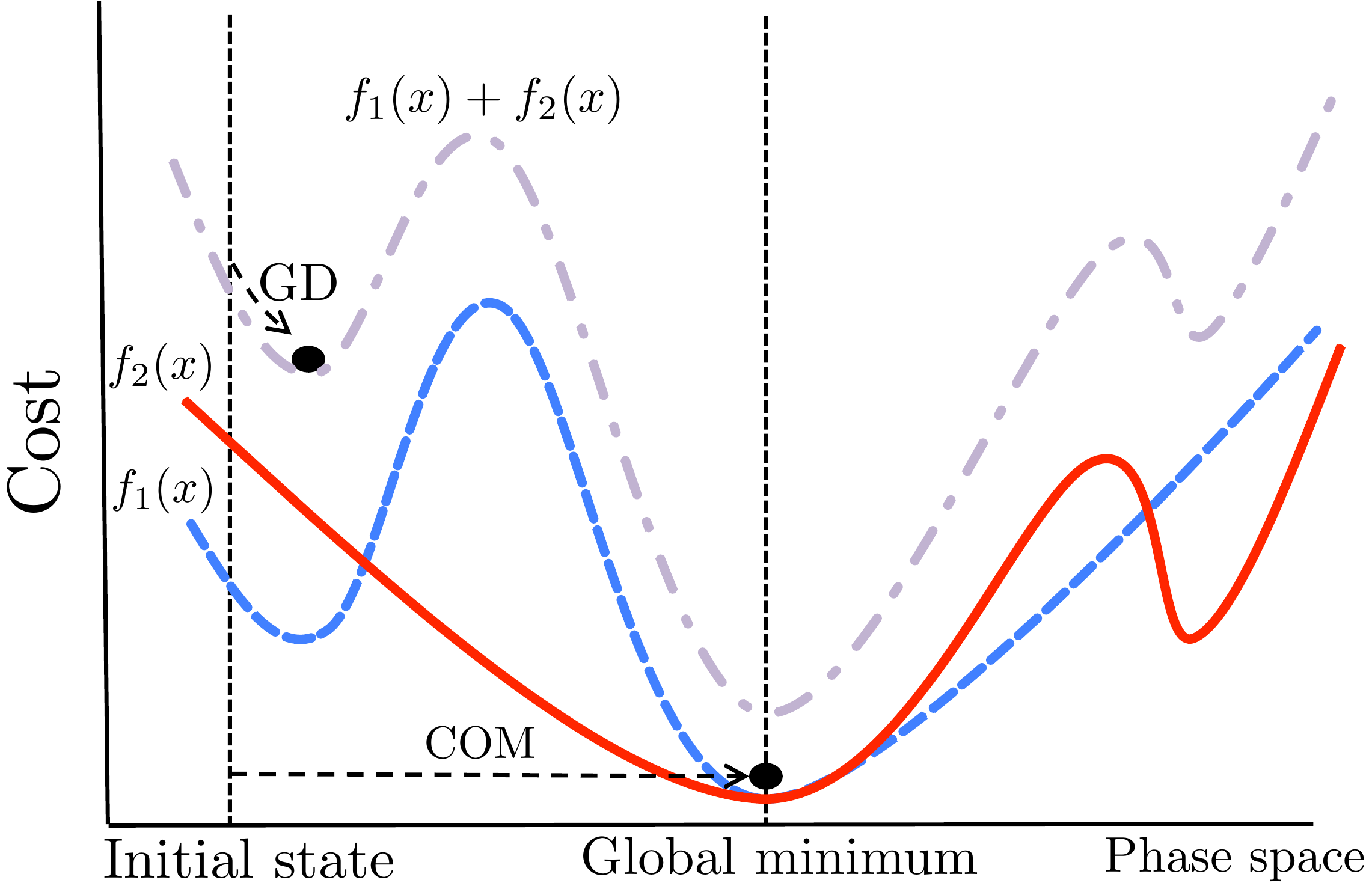}
  \caption{
    Schematic illustration of how the COM works for the case with $N=2$ and $D=1$: there are 3 intersections.
    In this case, from left or right intersection, traditional approach may not go to the global minimum but COM always goes.
    For example, We start the optimization from the initial state, represented by the left vertical dashed line.
    The standard gradient decent (GD) for $f_1(x) + f_2(x)$, (gray dashed curve), gets trapped at the nearest local minimum (left filled circle). On the other hand, in the COM, the optimization does {\em not} stop at the local minimum but keeps on moving to the right, since $df_2/dx < 0$ until arriving at the global minimum (right filled circle).
    By this way, the COM can overcome the barrier between local minima.
  }
  \label{Schematics_of_COM}
\end{figure}

By appropriately combining two or more cost functions that share the same global minimum but have different distributions of local minima, we expect that the cost of local minima is increased relative to the global minimum, and it may enhance the success rate of the optimization.
It should be noted, however, that the local minima can not be removed completely in general, even if the weights $\{\alpha_k\}$ are chosen carefully (see Fig.~\ref{Schematics_of_COM}).
This means that the gradient descent still gets trapped by one of the remaining local minima, and thus one has to introduce some {\em metaheuristics}, such as the simulated annealing~\cite{PannetierACC1990, DollSJ2008, TsujimotoAATT2018}.

\begin{algorithm}[H]
  \caption{Combined Optimization Method (COM)}
  \label{alg:COM}
  \begin{algorithmic}[1]
    \STATE Choose initial condition {\bf x} at random in the $D$-dimensional search space, set sign factors $\{s_i\}_{i=1}^D$ to either +1 or -1 at random, and set the initial learning rate $\Delta x$ appropriately, e.g., $\Delta x = 0.1$.
    \WHILE{$\Delta x$ $>$ threshold}
      \STATE Calculate the {\em generalized force} ${\bf F}=(F_1,\cdots,F_D)$ as
      \begin{align}
        F_i = \sqrt{\frac{1}{N} \sum_{k=1}^N \frac{1}{\|\nabla f_k \|^2} \bigg(\frac{\partial f_k}{\partial x_i} \bigg)^2} \qquad \text{for $i=1,\cdots,D$}.
        \label{eqn:com_force}
      \end{align}
      \STATE Determine the sign factors $\{s_i\}_{i=1}^D$ as
      \begin{align}
        s_i = \begin{cases}
          +1 \ & \text{if $(\partial f_k/\partial x_i) \geq 0$ for all $k$,} \\
          -1 \ & \text{if $(\partial f_k/\partial x_i) \leq 0$ for all $k$,} \\
          \text{unchanged} \ & \text{otherwise.}
        \end{cases}
        \label{eqn:com_sign}
      \end{align}
      \STATE Update ${\bf x}$ as
      \begin{align}
        x_i \leftarrow x_i - s_i F_i \Delta x \qquad \text{for $i=1,\cdots,D$.}
      \end{align}
      \STATE If all the sign factors change their signs, then
      \begin{align}
        \Delta x \leftarrow \Delta x / 2.
      \end{align}
    \ENDWHILE
  \end{algorithmic}
\end{algorithm}

In order to avoid being trapped by local minima, we introduce two essential modifications to the gradient descent.
Our algorithm, named ``Combined Optimization Method (COM),'' is summarized in Algorithm~\ref{alg:COM}.
First, we introduce a {\em generalized force} defined in equation~\ref{eqn:com_force}.
By definition, each component of the generalized force, $F_i$, is non-negative and becomes zero iff all the derivatives $\partial f_k / \partial x_i$ ($k=1,\cdots,N$) vanish.
This guarantees that the algorithm will stop only at the common minimum shared by all the cost functions.
Second, we introduce the {\em memory} to the direction of update. In equation~\ref{eqn:com_sign}, the sign factors $\{s_i\}$, which determines the direction of update, are chosen according to the {\em unanimity rule} for each component. When the votes are split, the sign factor of the previous step is inherited for that component. The last rule introduces the memory effect, or the {\em momentum}, in the optimization process, and helps the configuration ${\bf x}$ to escape from local minima of the individual cost functions.

It should be noted that the generalized force, {\bf F}, is always normalized to unity, i.e., $\| {\bf F} \| = 1$.
The learning rate in the present algorithm, $\Delta x$, has the same physical dimension as $x_i$, and thus one can choose the initial value of $\Delta x$ in a natural way by considering the physical property of the target problem as we will discuss later.
At step~6 in Algorithm~\ref{alg:COM}, the learning rate is halved when all the sign factors change their signs at the same time, which gradually decreases $\Delta x$ near the optimal point.

The proposed method reduces to the conventional gradient descent besides the learning rate, if we consider only one cost function, i.e., $N=1$. Although the COM is an extension of the gradient descent, its convergence behavior is qualitatively different due to the construction of the generalize force and the introduction of the memory effect.
Especially, the direction of optimization, i.e., the set of the sign factors $\{s_i\}$, is not determined by a particular cost function nor by the average of cost functions, but depends on the detailed landscape of all the cost functions via the unanimity rule (equation~\ref{eqn:com_sign}).
Nearby the common minimum, all the sign factors change their signs at every step, and as a result, a dumped oscillation occurs around the optimal point.
On the other hand, such an oscillation does not occur at a local minimum of each cost function, since the sign factors do not change at such a point (see Fig.~\ref{Schematics_of_COM}).

The above argument can be generalized to higher-dimensional cases ($D>1$).
Let us consider the gradient descent for a single cost function ($N=1$).
With a sufficiently small learning rate, the $D$-dimensional search space can be decomposed into a set of domains, each of which defines a ``basin of attraction'' of each global or local minimum of the cost function.
We define such a set of domains for each cost function also in the case of the joint optimization ($N>1$).
Then, we consider one of intersections of N domains.
By construction, $M$ of all $N$ functions have their minima in this intersection.
If $M$=$N$, the COM can not escape from the intersection. Indeed, the intersection containing the global minimum, which is shared by all the cost function, has exactly N minima. In other words, the COM gets trapped in the vicinity of the global minimum as we have expected.
If $M<N$, on the other hand, the COM can escape from the intersection, since at least one sign factor does not change its sign within the intersection.
This remarkable property of the COM enables us to reach the common minimum.

Indeed, the average number of intersections with exactly $N$ minima can be estimated as follows; Let $n_i$ the number of local minima in $i$-th cost function ($i=1 \cdots N$) and $n_{\rm s}$ the total number of intersections.
By construction $n_{\rm s} \ge n_i$ for $i=1 \cdots N$.
Note that the equality holds only when all the cost functions are identical with each other, and we can expect that $n_{\rm s}$ is much larger than $n_i$'s as we prepare the set of cost functions so that they have different distributions of local minima.
Then, since the probability that a randomly chosen intersection contains a local minimum of $i$-th cost function is given by $n_i/n_{\rm s}$, the average number of intersections with exactly $N$ minima is
\begin{align}
  \Big[ \prod_{i=1}^N \frac{n_i}{n_{\rm s}} \Big] \times n_{\rm s} =
  \frac{\prod_{i=1}^N n_i}{n_{\rm s}^{N-1}} \ll n_i \qquad \text{for $i=1 \cdots N$,}
\end{align}
that is, in average the number of such intersections always becomes much smaller the number of local minima of each cost function.
Thus, we expect that the COM process converges to the global minimum with high success rate, in comparison with the conventional joint optimization using a linear combination of cost functions.

\section*{Application to Crystal Structure Prediction}

In this section, we demonstrate the effectiveness of the COM, by applying it to an optimization problem that we encounter in materials science.
Here, we address the crystal structure prediction by simultaneously optimizing the theoretical potential energy and the degree of coincidence with experimental data, the same problem discussed by Tsujimoto {\it et al.}~\cite{TsujimotoAATT2018}.
As target materials, we consider Si diamond structure and two well-known polymorphs of SiO$_2$, low quartz and low cristobalite.
Theoretical structure prediction for SiO$_2$ is extremely difficult since there exist an enormous number of metastable structures.

We introduce two cost functions that have qualitatively different properties with one another.
The first cost function, $E({\bf x})$, is the theoretical potential energy of the crystal, which is a function of atomic position ${\bf x}$ and in general sensitive to the local structure between adjacent atoms.
For calculating the potential energy from the position of atoms, we adopt the classical force field, the Tersoff potential~\cite{Tersoff1989} and the Tsuneyuki potential~\cite{TsuneyukiTAM1988} for Si and SiO$_2$, respectively, by using the LAMMPS package~\cite{Plimpton1995} through the ASE~\cite{LarsenMBCCDFGHHHJJKKKKKLMMOPPRSSSTVVWZJ2017}.

The second cost function is defined using the powder X-ray diffraction pattern, which reflects the long-range periodicity of the crystal.
We calculate the diffraction intensity, $I_{\rm calc}(\theta;{\bf x})$, where $\theta$ is the diffraction angle, from the atomic position ${\bf x}$, and measure the degree of coincidence with the experimental diffraction intensity, $I_{\rm exp}(\theta)$. For the calculation of $I_{\rm calc}(\theta;{\bf x})$, we use the atomic scattering factors listed in International Tables for Crystallography~\cite{Prince2006}.
In the conventional crystal structure determination by using experimental X-ray diffraction data, the so-called $R$-value,
\begin{align}
  R({\bf x}) &= \frac{\int d\theta \, \| I_{\rm obs}(\theta) - I_{\rm calc}(\theta;{\bf x}) \|}{\int d\theta \, I_{\rm obs}(\theta)},
  \label{R_p}
\end{align}
has been used widely~\cite{HarrisT1996,PutzSJ1999}.
This $R$-value is, however, sensitive to the noise and/or the background of the experimental data, and thus it does not necessarily give the global minimum even at the correct crystal structure.
In the present work, we adopt a cost function that does not use the intensity information of the X-ray diffraction peaks~\cite{TsujimotoAATT2018}, which is defined by
\begin{align}
  D({\bf x}) = 1 - \lambda({\bf x}),
\end{align}
where $\lambda({\bf x})$ is called ``crystallinity'' calculated using only the {\em positions} of the experimental X-ray diffraction peaks, i.e.,
\begin{align}
  \lambda({\bf x}) = \frac{\displaystyle \sum_{\theta_{\rm obs}} \int_{\theta_{\rm obs}-\delta}^{\theta_{\rm obs}+\delta} I_{\rm calc}(\theta;{\bf x})\,d\theta}{\displaystyle \int I_{\rm calc}(\theta;{\bf x})\,d\theta}.
\end{align}
Here, $I_{\rm calc}(\theta;{\bf x})$ denotes the intensity of the calculated diffraction pattern for atomic position ${\bf x}$, $\theta_{\rm obs}$ means the peak positions in the reference (e.g., experimentally observed) diffraction pattern, and $\delta$ is the diffraction angle resolution.
In our calculations the diffraction angle resolution $\delta$ is set to 0.1$^\circ$, and the X-ray wavelength is set to 0.1541~nm, which corresponds to the Cu~$K\alpha$ radiation.
The range of integration in the numerator is restricted to the vicinity of the peaks observed in the experiment, and that of the denominator is the whole measured angle.
By definition, $0 \le D({\bf x}) \le 1$; for the correct crystal structure, where all the calculated peak positions coincide with those observed in the experiment, we have $D({\bf x}) = 0$, otherwise $D({\bf x})$ becomes nonzero.

The global minimum of $D({\bf x})$ is more robust against the noise in the experimental data.
It should be noted, however, that other non-trivial global minima with $D({\bf x})=1$ may appear, since it does not take into account the intensity.
Although this fact makes more difficult to determine the correct structure by using the crystallinity alone relative to the $R$-value, it has been shown that the crystal structure prediction becomes much more reliable by combining the crystallinity with the the theoretical potential calculation~\cite{TsujimotoAATT2018}. In what follows, we demonstrate the COM is able to accelerate further the structure determination by 
using the property that the theoretical potential energy and the crystallinity share the identical target structure as the their global optimum.

\subsection*{Si diamond structure}
We use a $2 \times 2 \times 2$ cubic supercell of 64 Si atoms. The linear size of the supercell is fixed to $2 \times 5.431$~\AA, which is used as the unit of ${\bf x}$ and $\Delta x$. The periodic boundary conditions are imposed. 
The initial position of atoms is sampled uniformly at random in the supercell.
For the definition of the crystallinity, we use only three peaks located between 20$^\circ$ and 60$^\circ$.

\begin{figure}[tb]
  \centering
    \includegraphics[clip, width=\linewidth]{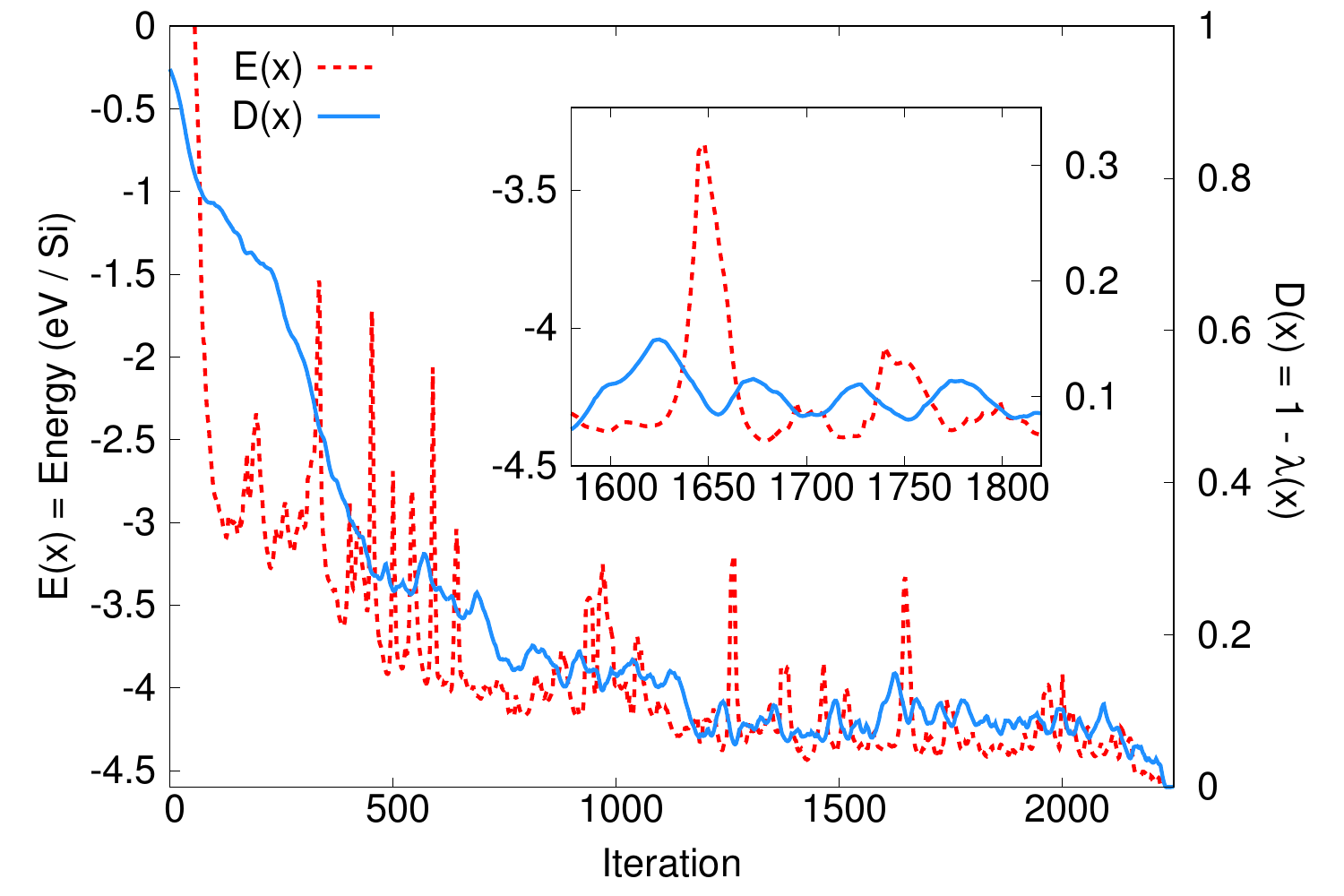}
    \caption{Change in the cost functions, $E({\bf x})$ (red dotted line) and $D({\bf x})$ (blue solid line) of the Si system with $\Delta x = 0.01$. In the inset, the data between step 1600 and 1800 is enlarged, where an {\em anti-phase} oscillation of $E({\bf x})$ and $D({\bf x})$ is observed. We observe that the sum of the two cost functions {\em increases}, for example, during step from 1625 to 1650, and as a result a barrier between local minima is overcome. This can be archived by the memory effect introduced in the COM.}
    \label{Si_both}
\end{figure}

First, we show a typical change in the cost functions, $E({\bf x})$ and $D({\bf x})$, for $\Delta x = 0.01$ during the COM optimization process in Fig.~\ref{Si_both}.
In the early stage of the optimization (before step $\simeq 1000$), we observe that $D({\bf x})$ decreases almost monotonically. As for $E({\bf x})$, although it intermittently exhibits a sharp spike structures, both of the values at spikes and at valleys decrease gradually.
This early stage behavior is similar to that of the standard steepest descent.
In the middle stage (step $1000 \sim 2000$), both cost functions show oscillating behavior.
More precisely, $E({\bf x})$ and $D({\bf x})$ show an anti-phase oscillation with each other.
This indicates that the configuration is 
repeatedly trapped at some local minimum of one of the two cost functions and escapes from it one after another.
This behavior can not happen in the standard steepest descent.
The conventional joint optimization using the linear combination of cost functions is not able to escape from local minima without the help of metaheuristics, while the COM can get out of such structures.
This is because the sign factors do not change when the gradients of the cost functions conflict with each other.
Finally, after step $\simeq 2000$, they start to decrease again and reach the minimum values successfully, which correspond to the correct Si diamond structure.

\begin{figure}[tb]
  \centering
    \includegraphics[clip, width=\linewidth]{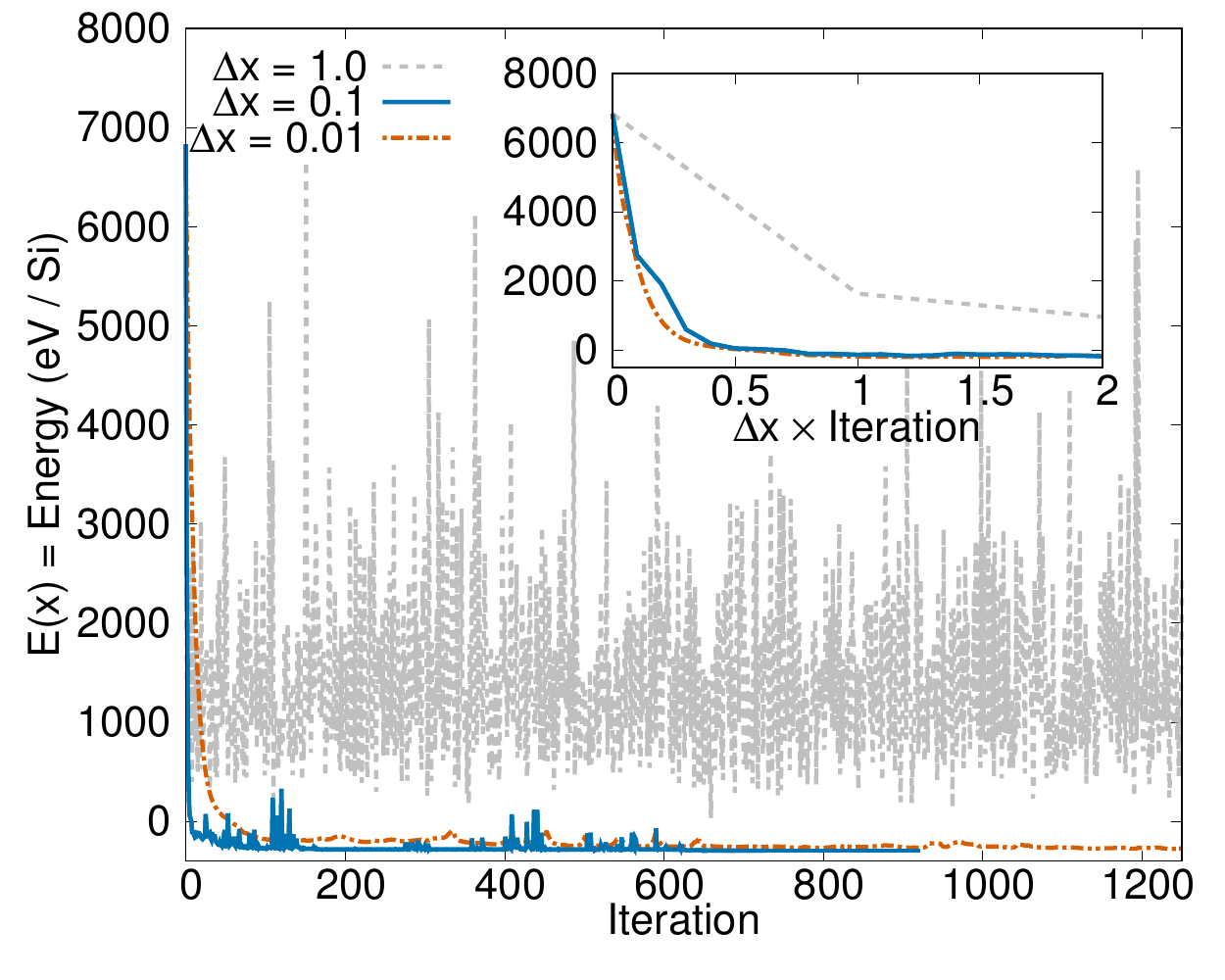}
    \caption{Change in the potential energy $E({\bf x})$ of the Si system with $\Delta x=1.0$ (gray dashed line), 0.1 (solid blue line), and 0.01 (red dashed-dotted line). With $\Delta x = 0.1$ and $0.01$, the potential energy reaches the optimal value after the optimization, while with $\Delta x = 1.0$ the energy continues to fluctuate and does not converge. In the inset, the horizontal axis is multiplied by $\Delta x$. In the cases with $\Delta x = 0.1$ and 0.01, the potential energy shows a similar initial relaxation curve, while with $\Delta x = 1.0$ the potential energy exhibits different behavior and stays at higher values than the other cases.}
    \label{Si_energy}
\end{figure}

Next, we discuss the $\Delta x$-dependence of the performance of the COM.
In Fig.~\ref{Si_energy}, we show the change in the potential energy $E({\bf x})$ for $\Delta x = 0.01$, $0.1$, and $1.0$ starting from the same initial configuration.
It is observed that when the initial learning rate is sufficiently small, e.g., $\Delta x = 0.01$ or 0.1, the potential energy exhibits a quick relaxation at the very early stage of the optimization, and successfully reaches the correct structure.
If the learning rate is too large, i.e., $\Delta x = 1.0$, on the other hand, the potential energy fluctuates strongly and does not converge even after 1000 iterations.
In the inset of Fig.~\ref{Si_energy}, we show the same data with the horizontal axis multiplied by $\Delta x$. We observe the similar behavior in the initial relaxation for $\Delta x = 0.01$ and 0.1.
This means that the optimization trajectories in the $(3 \times 64)$-dimensional search space are almost identical in these two cases, though the former requires 10 times more iterations to proceed to the same distance.
With $\Delta x = 1.0$, the initial relaxation follows a different trajectory from the other cases.
This implies that $\Delta x = 1.0$ is too large in comparison with the typical length scale of the potential energy landscape, and thus smaller $\Delta x$ (0.1 or 0.01) seems to be more appropriate.

Let us see the $\Delta x$-dependence of the success rate of the optimization.
We prepare 100 different initial configurations where the position of atoms is chosen uniformly at random in the supercell.
We stop the COM process when $\Delta x$ becomes smaller than 0.005, or the number of iterations exceeds some threshold (5000 for $\Delta x = 1.0$ and $0.1$, and 50000 for $\Delta x = 0.01$).
After the optimization by the COM, we additionally perform the gradient descent optimization on the potential energy for 200 steps.
The success rate of the optimization strongly depends on the initial learning rate. For $\Delta x = 1.0$, we observe that no samples converge to the correct Si diamond structure.
For $\Delta x = 0.1$ and 0.01, about 90 samples among 100 trials reach the correct structure successfully.
We point out that although the cases with $\Delta x = 0.1$ and 0.01 exhibit similar success rates, the former converges with smaller number of iterations than the latter. Thus, we conclude $\Delta x = 0.1$ for the optimal initial learning rate.

In the present Si diamond case
with $\Delta x = 0.1$, since the generalized force (equation~\ref{eqn:com_force}) is normalized to unity and we have $3 \times 64 = 192$ coordinate variables in total for 64 Si atoms, the average displacement in each component of ${\bf x}$ by one iteration step is evaluated approximately as $2 \times 5.431 \times 0.1 / \sqrt{192} \approx 0.078~\AA$.
This average displacement is smaller enough than the distance between Si atoms in the Si diamond structure, $5.431 \times \sqrt{3}/4 \approx 2.35~\AA$.
With $\Delta x = 1.0$, on the other hand, the average displacement ($\approx 0.78~\AA$) is the same order as the bond length between Si atoms, and thus is too large to capture the structure of the potential energy landscape.
This rough estimation for the appropriate value for $\Delta x$ is consistent with the results of the numerical test presented above.

\subsection*{SiO$_2$ polymorphs}

\begin{figure}[tb]
  \centering
    \includegraphics[clip, width=\linewidth]{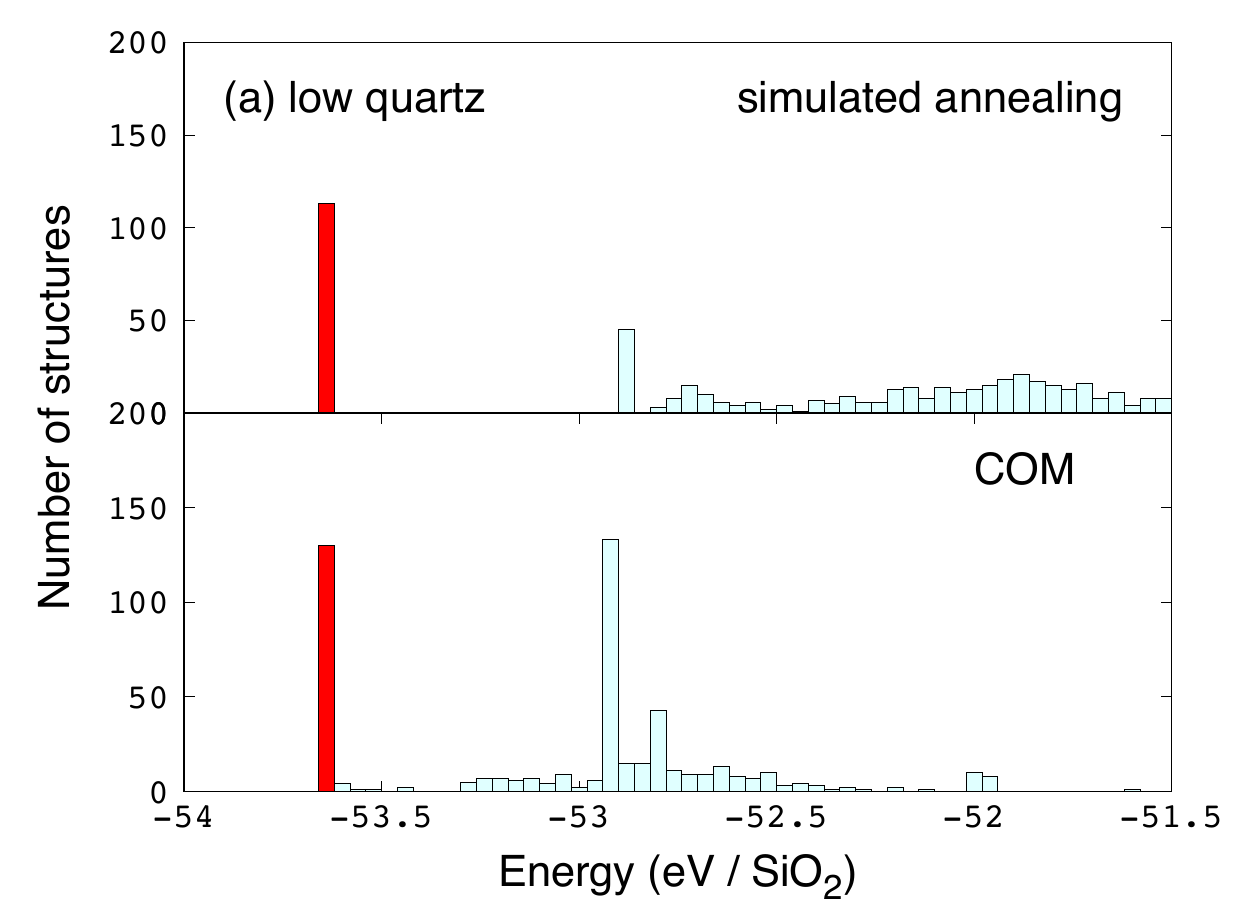}
    \includegraphics[clip, width=\linewidth]{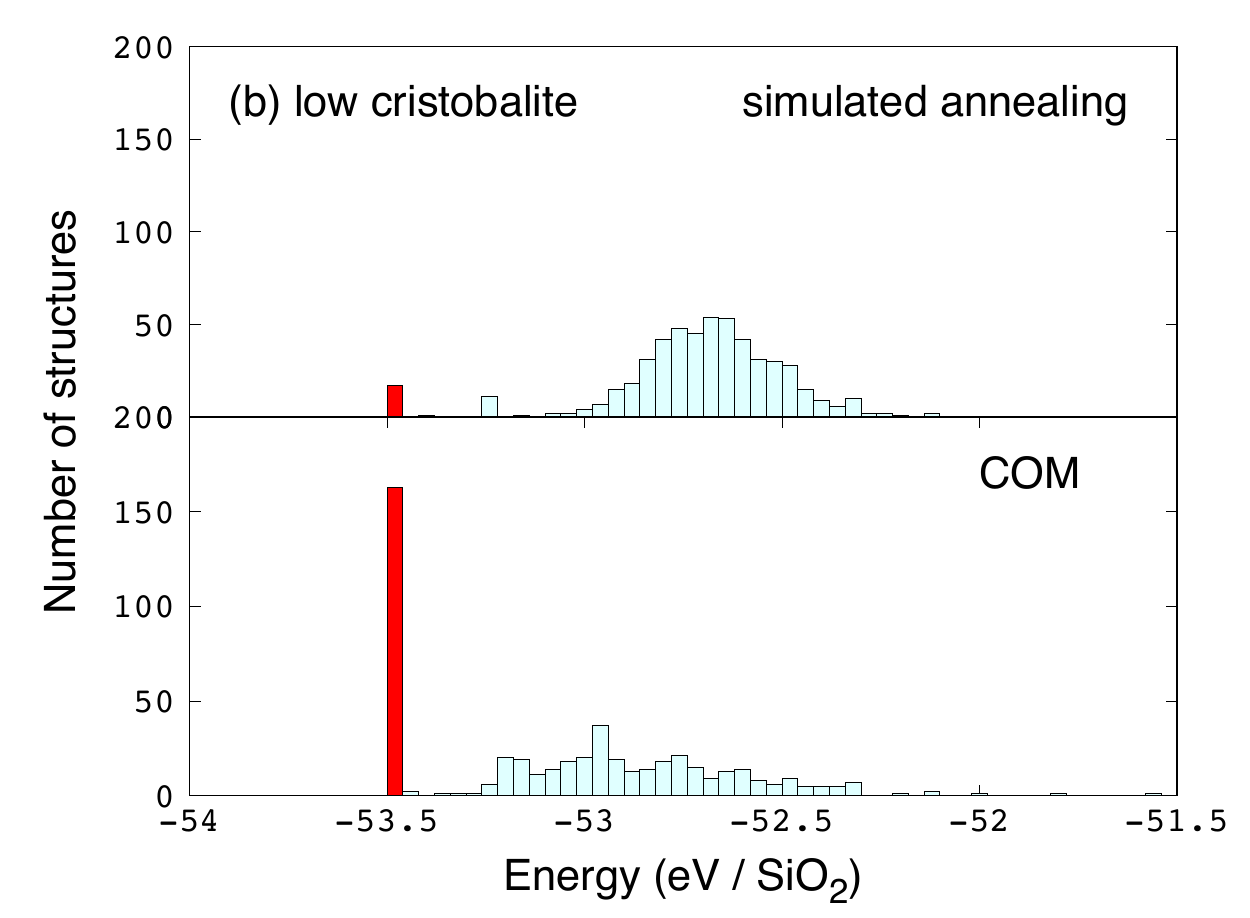}
    \caption{Histogram of the potential energy after the optimization for 500 trials for (a) low quartz and (b) low cristobalite. The results of the present method is shown in the lower half in each graph. In the upper half in each graph, for comparison, the previous results by Tujimoto {\em et al} is presented~\cite{TsujimotoAATT2018}. The red bars denote the samples that reached the correct structures which have $E=-53.65$~eV/SiO$_2$ and $-53.48$~eV/SiO$_2$ for low quartz and low cristobalite, respectively. Note that the total number of iterations in the previous study is about five times longer than the present simulation.}
    \label{SiO_histograms}
\end{figure}

Next, we show the results for low quartz and low cristobalite. We consider $2 \times 2 \times 1$ supercells containing 36 and 48 atoms for low quartz and low cristobalite, respectively. The lattice parameters are fixed to the same as those used in Ref.~\cite{TsujimotoAATT2018}.
For the definition of the crystallinity, we use seven and six peaks located between 20$^\circ$ and 45$^\circ$ for low quartz and low cristobalite, respectively.
In Fig.~\ref{SiO_histograms}, we show the histograms of the potential energy after the optimization. The initial learning rate is set as $\Delta x = 0.1$ in both cases. 
The COM process is terminated when $\Delta x$ becomes smaller than 0.005, or the number of iterations exceeds 1000.

For comparison, in Fig.~\ref{SiO_histograms}, we also present the previous results obtained by the molecular dynamics with simulated annealing for $E({\bf x}) + \alpha N D({\bf x})$, where $N$ is the number of the atoms in the supercell and the weight $\alpha$ is set to the optimal value $\alpha/k_{\rm B} \simeq 4.5 \times 10^4$~K~\cite{TsujimotoAATT2018}. Here, $k_{\rm B}$ is the Boltzmann constant. The integration time step of the molecular dynamics simulation is 1~fs, and the temperature is kept at 500~K for the first 4500 steps and then gradually quenched to 0~K in 700 steps.

As shown in Fig.~\ref{SiO_histograms}, by using the COM, 130 and 163 samples among 500 trials converge to the correct structure for low quartz and low cristobalite, respectively.
These success rates are significantly higher than the previous simulated-annealing results presented in the upper half in each graph, especially for low cristobalite.
It should be pointed out that the total number of iterations for the simulated annealing~\cite{TsujimotoAATT2018} was 5200, which is more than five times longer than the present simulation.
For the low quartz case, there exists another large peak around $E \simeq -52.9$~eV/SiO$_2$.
This structure is also a common local minimum, but is an artifact due to the fixed lattice parameters.
Indeed, once we further optimize the cell size as a free parameter starting from this local minimum structure, it converges to the correct low quartz structure rapidly, and the total success rate increases by about two times.

We have also applied the COM to coesite, another polymorph of SiO$_2$, but found a lower success rate than the simulated annealing.
This failure might come from the fact that there are a number of common local optimal structures of the two cost functions in the case of coesite.
Although these structures have significantly higher energies than the correct coesite structure, the COM is easily trapped to these local minima since they are common to the two cost functions.
In such a case, the COM would not be very useful for searching the global optimal structure, since the COM is just an extension of the gradient descent.

\section*{Conclusion and Discussion}

In this paper, we presented a new optimization method, the Combined Optimization Method (COM), for the joint optimization of multiple cost functions that share a common global minimum.
In the COM, we have introduced two essential ingredients, the generalized force and the memory (Algorithm~\ref{alg:COM});
They enable the configuration to escape from local minima of each cost function and explore in wider search space, and as a result guarantee that the optimization will stop at the common minimum shared by all the cost functions.

We demonstrated the effectiveness of our method by applying it to the crystal structure prediction in materials science.
As cost functions, we used the theoretical potential energy of the crystal and the crystallinity, which characterizes the agreement with the theoretical and experimental X-ray diffraction patterns.
The former is sensitive to the local structure between neighboring atoms, while the latter reflects the long-range periodicity of the crystal.
As a result, it is expected that they have qualitatively different distributions of local minima besides the global minimum, which corresponds to the correct crystal structure.
We have applied our method to Si diamond structure, low crystal, and low cristobalite, and observed that the correct target structures can be found with significantly higher success rate than the previous work.

In the simulation performed in the present paper, we have fixed the cell parameters so that the potential energy has a global minimum exactly for each target crystal structure, and in addition, we have used the theoretically exact peak positions of each target structure in the definition of the crystallinity.
Therefore, in the present examples, it is guaranteed that the global optimal points of the two cost functions strictly coincide with each other.
Even though the situation assumed in the present work sounds too idealized, and in reality the optimal positions do not necessarily coincide strictly due to experimental and/or numerical errors or approximations, the COM should still find the target structure with high success rate with a help of some traditional local optimization method at the final stage as long as the two minima are involved in the same intersection of basins of attraction.

The COM has one tuning parameter, the initial learning rate, $\Delta x$.
If the initial value of $\Delta x$ is too large, it does not decrease and strong fluctuation persists asymptotically.
If $\Delta x$ is too small, on the other hand, it requires more iterations to explore the whole search space.
In the present implementation of the COM, the learning rate is halved when all the sign factors change their signs at the same time (step~6 in Algorithm~\ref{alg:COM}).
From our experience in the application to the crystal structure prediction, this rule seems to be too strict: the reduction of $\Delta x$ rarely happens during the optimization process in higher dimensional space.
This causes asymptotic strong fluctuation with too large $\Delta x$ as well as the need of (short) gradient descent iterations at the final stage of the optimization.
Although $\Delta x$ has the same physical dimension as $x_i$, and thus one can choose the initial value of $\Delta x$ in a natural way by considering the physical property of the target problem, a better choice of the initial value of $\Delta x$ as well as a more robust criterion for the reduction of the learning rate is a subject of future studies.

\section*{Acknowledgements}

This work was supported by Elements Strategy Initiative to Form Core Research Center in Japan, Japan Society for the Promotion of Science KAKENHI Grant No. 17H02930, the ``Materials Research by Information Integration'' Initiative (MI2I) Project of the Support Program for Starting Up Innovation Hub from Japan Science and Technology Agency (JST), and Ministry of Education, Culture, Sports, Science and Technology (MEXT) of Japan as a social and scientific priority issue (Creation of new functional devices and high-performance materials to support next-generation industries; CDMSI) to be tackled by using post-K computer.
D.A. and N.T. are supported by the Japan Society for the Promotion of Science through the Program for Leading Graduate Schools (MERIT).

\bibliography{draft}

\end{document}